\documentclass[cameraready]{Interspeech}

\title{Discrete Token Modeling for Multi-Stem Music Source Separation with Language Models}

\author[affiliation={1}]{Pengbo}{Lyu}
\author[affiliation={1}, correspondingauthor]{Xiangyu}{Zhao}
\author[affiliation={1}]{Chengwei}{Liu}
\author[affiliation={1}]{Haoyin}{Yan}
\author[affiliation={1}]{Xiaotao}{Liang}
\author[affiliation={1}]{Hongyu}{Wang}
\author[affiliation={1}]{Shaofei}{Xue}

\address{
    $^1$ Qwen Applications Business Group of Alibaba, China
}

\email{\{lyupengbo.lpb, zhaoxiangyu.zyx, liuchengwei.lcw, yanhaoyin.yhy, xiaotao.lxt, hongxin.why\}@alibaba-inc.com, mullerxue@126.com}


\usepackage{comment}
\usepackage{booktabs}
\usepackage{multirow}
\usepackage{array} 
\usepackage{amsmath}
\usepackage{amssymb}

\begin{document}

\maketitle

\begin{abstract}
We propose a generative framework for multi-track music source separation (MSS) that reformulates the task as conditional discrete token generation. Unlike conventional approaches that directly estimate continuous signals in the time or frequency domain, our method combines a Conformer-based conditional encoder, a dual-path neural audio codec (HCodec), and a decoder-only language model to autoregressively generate audio tokens for four target tracks. The generated tokens are decoded back to waveforms through the codec decoder. Evaluation on the MUSDB18-HQ benchmark shows that our generative approach achieves perceptual quality approaching state-of-the-art discriminative methods, while attaining the highest NISQA score on the vocals track. Ablation studies confirm the effectiveness of the learnable Conformer encoder and the benefit of sequential cross-track generation. 
\footnote{Demo page: \url{https://anonymous.4open.science/w/mss-demo-page-2F80/}. }

\end{abstract}

\section{Introduction}

Music Source Separation (MSS) aims to decompose a mixture of audio signals into individual sources, such as vocals, drums, bass, and other instruments. This task is central to applications such as music remixing, transcription, karaoke generation, and enhancing the experience of hearing-impaired individuals. 
Current mainstream MSS methods primarily explore frequency-domain \cite{uhlich2017improving, tong2024scnet} or time-frequency fusion \cite{defossez2021hybrid, kim2021kuielab} approaches.
Despite significant progress \cite{mitsufuji2022music, Fabbro_2024}, MSS remains challenging due to the complexity and high dimensionality of audio signals, which vary widely across source types and musical contexts.

Many deep learning approaches to MSS, including convolutional neural networks \cite{chandna2017monoaural}, recurrent neural networks \cite{luo2023music}, and U-Net-based models \cite{defossez2019music}, operate in the frequency domain by processing spectrogram features derived from the Short-Time Fourier Transform (STFT). More recently, transformer-based architectures such as Hybrid Transformer Demucs \cite{rouard2023hybrid} and BS-RoFormer \cite{lu2024music} have achieved state-of-the-art results by modeling global dependencies. However, these discriminative methods still estimate continuous spectral masks or waveforms, inheriting the limitations of regression-based formulations.

The emergence of neural audio codecs such as SoundStream \cite{zeghidour2021soundstream} and EnCodec \cite{defossez2022high} has enabled continuous audio to be encoded into discrete token sequences, facilitating research in audio generation and discrete audio modeling \cite{borsos2023audiolm, chen2025neural}. TokenSplit \cite{erdogan2023tokensplit} explored the use of discrete token representations for speech separation with a sequence-to-sequence Transformer architecture, demonstrating effective separation and transcription in the speech domain. More recently, TSELM \cite{tang2025tselm} proposed a target speaker extraction framework that leverages discrete tokens and language models to integrate target speaker information for extraction tasks. UniSep \cite{wang2025unisep} further extended the use of language models to universal target audio separation over arbitrary mixtures of audio types. However, these methods are primarily designed to extract a single target source from a mixture (e.g., a specific speaker or target sound) rather than directly output multiple separated tracks simultaneously.

In this paper, we propose an MSS framework that combines a neural audio codec (HCodec \cite{liu2025unitok, liu2025quarkaudio}) with a decoder-only language model. Unlike discriminative methods that rely on spectrogram masking, our approach treats MSS as a discrete token generation task. We leverage residual vector quantization (RVQ) to represent each target track as interleaved acoustic and semantic tokens, which are autoregressively generated by the language model conditioned on the mixture. Our method operates at 48 kHz and autoregressively generates four target tracks sequentially in a single run, which facilitates modeling cross-track dependencies.

Our contributions are as follows: (1) We present a decoder-only autoregressive LM-based framework for high-resolution, multi-track music source separation. (2) We evaluate the framework on the MUSDB18-HQ \cite{musdb18-hq} benchmark and demonstrate that this generative approach achieves perceptual quality competitive with state-of-the-art discriminative methods, validating the effectiveness of discrete token modeling for multi-track music source separation.

\begin{figure*}[t]
  \centering
  \includegraphics[width=0.75\textwidth]{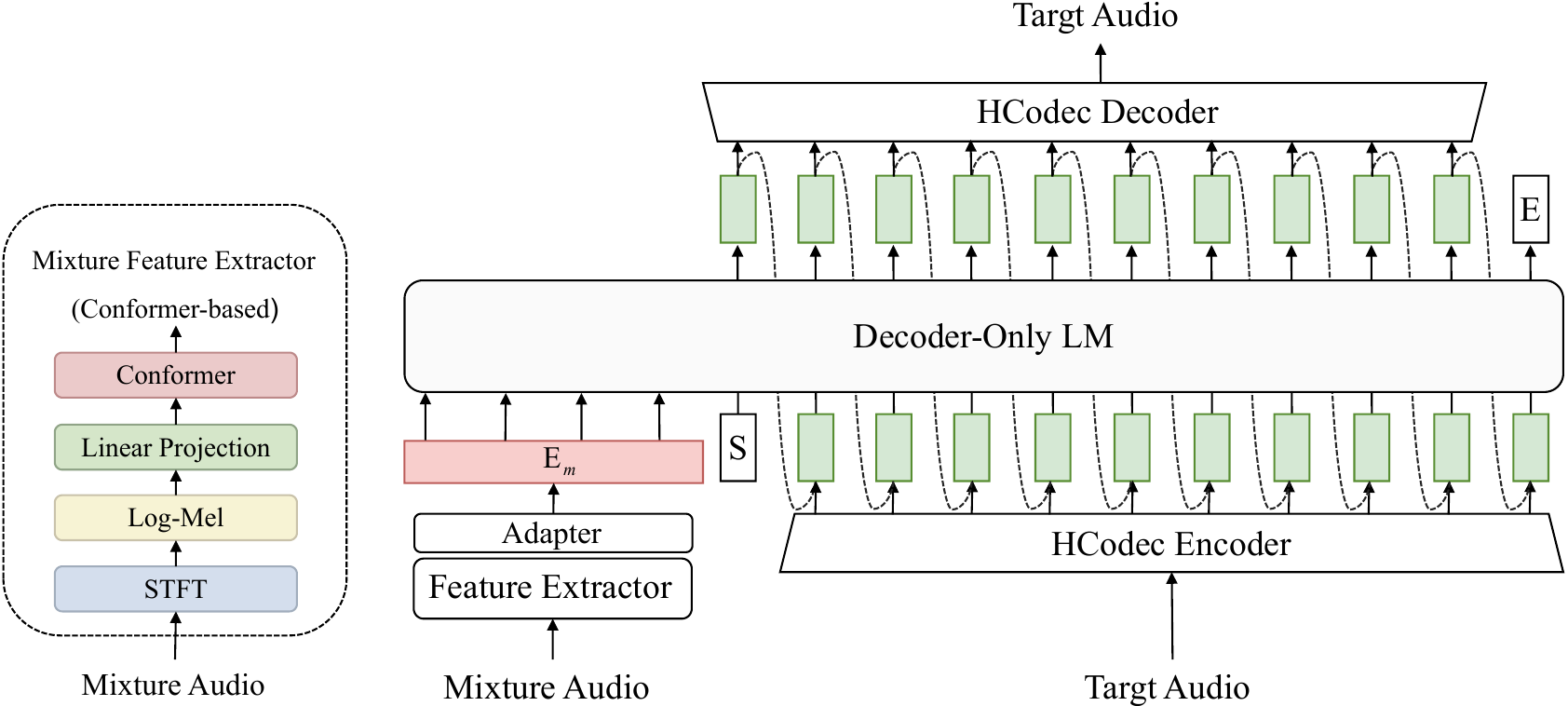}
  \caption{Architecture of the proposed token-based generative source separation model. The Mixture Embedding $E_m$ is conditioned by the Decoder-Only LM, which autoregressively generates discrete audio tokens.}
  \label{fig:LM_Structure}
\end{figure*}

\section{Method}

\subsection{Overall Architecture}

Given a mixture waveform $x_{\mathrm{mix}} \in \mathbb{R}^T$ sampled at 48 kHz, the framework predicts discrete token sequences for four target tracks $\{x_s\}_{s \in \mathcal{S}}$, where $\mathcal{S} = \{\mathrm{vocals},\, \mathrm{drums},\, \mathrm{bass},\, \mathrm{other}\}$, and reconstructs the separated waveforms via a neural codec decoder.

As illustrated in Figure~\ref{fig:LM_Structure}, the proposed framework consists of three components: (1) a conditional encoder that extracts continuous feature representations from the mixture waveform, serving as the conditioning input for the decoder-only language model; (2) a dual-path neural audio codec (HCodec \cite{liu2025unitok, liu2025quarkaudio}) that converts each target track into acoustic and semantic discrete token sequences through residual vector quantization (RVQ), and is also responsible for waveform reconstruction; and (3) a decoder-only language model that autoregressively generates multi-track token sequences conditioned on the mixture representation. The design of each component is detailed in Sections 2.2–2.4, and the training and inference procedures are described in Section 2.5.

\subsection{Conditional Feature Extraction}
The conditional encoder extracts compact features from the mixture waveform to condition the language model. We first compute a 120-band log-Mel spectrogram from the STFT of $x_{\mathrm{mix}}$ (FFT size 2048, hop length 960), yielding features $\mathbf{M} \in \mathbb{R}^{T' \times D_{\mathrm{mel}}}$. These are processed by a Conformer encoder with rotary positional embeddings and depthwise convolutions, producing $\mathbf{F}_{\mathrm{mix}} \in \mathbb{R}^{T' \times D}$, which is projected to the language model's hidden dimension through a linear adapter layer.

\subsection{Discrete Audio Tokenization}

We adopt HCodec \cite{liu2025unitok,liu2025quarkaudio} as the neural audio codec to convert continuous waveforms into discrete token sequences. HCodec employs a dual-path architecture with an acoustic encoder and a semantic encoder, both producing features at a frame rate of 12.5 Hz.

The acoustic path encodes STFT-derived spectral features via a ConvNeXt-Transformer network, while the semantic path extracts representations from a frozen HuBERT model \cite{hsu2021hubert} through a convolutional adapter. Both paths are aligned to the same frame rate.

Each path is independently quantized using a 16-layer RVQ with a codebook size of 1024, yielding acoustic tokens $\mathbf{c}^{\mathrm{a}}$ and semantic tokens $\mathbf{c}^{\mathrm{s}}$, each spanning $T'$ codec frames. To enable the language model to jointly capture acoustic and semantic information, we interleave the two token streams along the temporal axis:

\begin{equation}
\mathbf{c} = [c^{\mathrm{a}}_{0},\, c^{\mathrm{s}}_{0},\, c^{\mathrm{a}}_{1},\, c^{\mathrm{s}}_{1},\, \ldots,\, c^{\mathrm{a}}_{T'-1},\, c^{\mathrm{s}}_{T'-1}]
\end{equation}

For notational simplicity, we omit the RVQ layer index. In practice, each $c^{\mathrm{a}}_t$ and $c^{\mathrm{s}}_t$ corresponds to 16 codebook indices across the RVQ layers. The codec parameters are frozen throughout training.

\subsection{Autoregressive Token Generation}

To model the sequential dependencies among discrete tokens, we build the generation module upon a decoder-only Transformer with the LLaMA architecture \cite{touvron2023llama}. The model learns to predict the conditional distribution of each target track's token sequence given the encoded mixture representation.

The input sequence to the language model is organized as:

\begin{equation}
    x = \bigl[\langle \mathrm{mix} \rangle,\, \mathbf{F}_{\mathrm{mix}},\, \mathrm{S},\, \mathbf{c}^{(1)},\, \mathrm{S},\, \mathbf{c}^{(2)},\, \mathrm{S},\, \mathbf{c}^{(3)},\, \mathrm{S},\, \mathbf{c}^{(4)}\bigr]
    \label{eq:input_seq}
\end{equation}

where $\langle \mathrm{mix} \rangle$ marks the start of the mixture conditioning features $\mathbf{F}_{\mathrm{mix}}$, which are projected to the same hidden dimension as the token embeddings and serve as a continuous prefix. $\mathrm{S}$ is a shared start token preceding each track, and $\mathbf{c}^{(k)}$ is the interleaved token sequence of the $k$-th track. The four tracks follow a fixed order: vocals, drums, bass, and other. The corresponding output sequence is:

\begin{equation}
    y = \left[ \mathbf{c}^{(1)}, \mathrm{E}, \mathbf{c}^{(2)}, \mathrm{E}, \mathbf{c}^{(3)}, \mathrm{E}, \mathbf{c}^{(4)}, \mathrm{E} \right]
    \label{eq:output_seq}
\end{equation}

where $ \mathrm{E} $ is a shared end token marking the boundary of each track.

The model parameters $\theta$ are optimized by minimizing the negative log-likelihood:

\begin{equation}
    \mathcal{L} = - \sum_{t=1}^{L} \log P\left( y_t \mid \langle \mathrm{mix} \rangle, \mathbf{F}_{\mathrm{mix}}, y_{<t}; \theta \right)
    \label{eq:nll_loss}
\end{equation}

where $L$ is the total output length across all four tracks.

\subsection{Training and Inference Strategies}

During training, each track is tokenized by the frozen HCodec encoder into interleaved acoustic-semantic sequences. The four sequences are concatenated and prepended with the mixture features to form a single training sequence. The model is trained with teacher forcing under causal masking, with a weighted loss that prioritizes the first RVQ layer.

At inference, the mixture is first encoded into $\mathbf{F}_{\mathrm{mix}}$, which is fed as a prefix into the language model to produce initial key-value states. The four tracks are generated autoregressively in the same fixed order, with the cache preserved across tracks to leverage cross-track context. The model generates tokens autoregressively, using the end token $\mathrm{E}$ as a delimiter between tracks, and terminates when the total output reaches the predefined sequence length. The generated tokens are then de-interleaved and decoded by HCodec.

\begin{figure*}[t]
  \centering
  \includegraphics[width=0.8\textwidth]{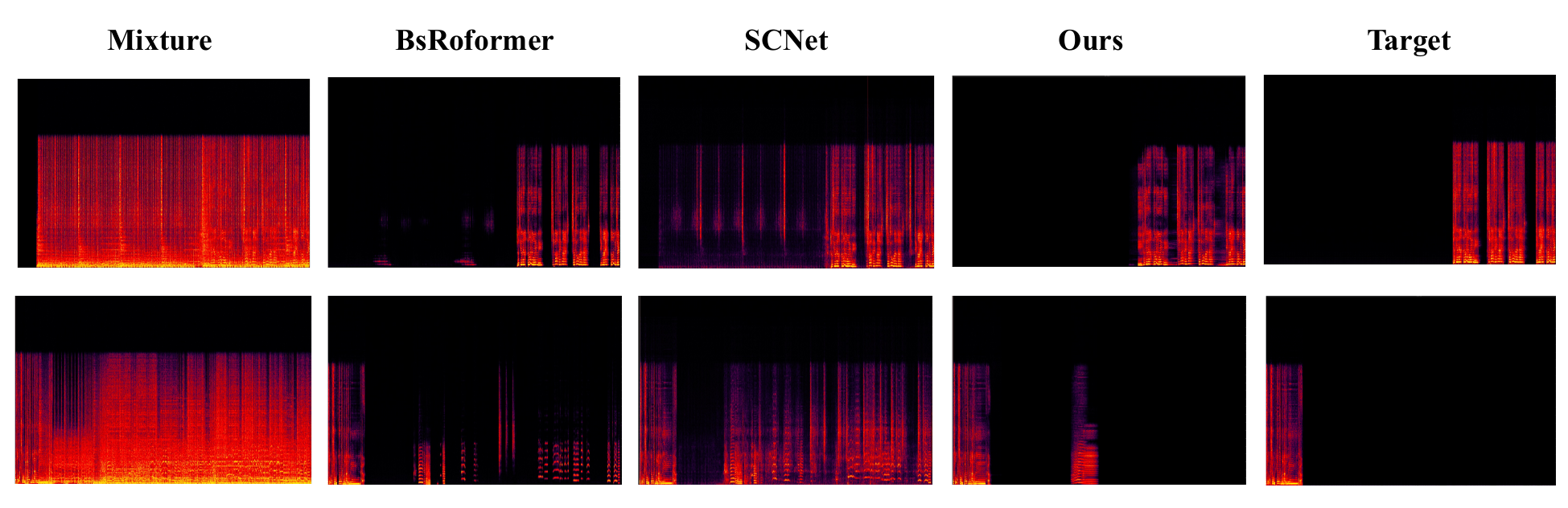}
  \caption{Qualitative spectrogram visualization of source separation results for the vocals track.}
  \label{fig:compare}
\end{figure*}

\section{Experimental Setup}
\subsection{Dataset and Data Augmentation}

We train the proposed model on a large-scale internal music dataset with approximately 23{,}000 hours of 44.1 kHz audio. The dataset includes songs, audiobooks, and instrumental tracks. Since the original recordings do not provide isolated track annotations, we apply BS-RoFormer \cite{lu2024music}, a state-of-the-art music source separation model, to the dataset. This produces pseudo ground-truth labels for the four target tracks: vocals, drums, bass, and other instruments.

To obtain high-quality training segments, we run voice activity detection (VAD) using the Silero VAD model \cite{silero-vad} on all vocal tracks, merging adjacent segments and discarding those shorter than 2.0 seconds.

Since the pseudo ground-truth is produced by BS-RoFormer, the quality of the training targets is inherently bounded by the teacher model. We nevertheless include BS-RoFormer as a baseline because the comparison reveals how much of the teacher's separation ability the generative framework can retain under a fundamentally different modeling paradigm, and whether discrete token generation introduces additional degradation beyond the teacher's own limitations.

During training, we apply online data augmentation including random loudness scaling per track ($[0.5, 1.5]$), polarity inversion (10\% probability), and a seven-band parametric equalizer with track-dependent gains.

We evaluate the proposed method on the MUSDB18-HQ test set \cite{musdb18-hq}, which consists of 50 full-length songs with professionally produced tracks (vocals, drums, bass, and other).

\subsection{Model and Training Configuration}

The conditional encoder is an 8-layer Conformer with 12 attention heads and a depthwise convolution kernel size of 31. The LLaMA-based decoder-only backbone consists of 16 layers, each with 16 attention heads and a hidden dimension of 2048, with dropout set to 0.1.

We train the model for 35 epochs on 8 NVIDIA A100 GPUs (80 GB). The pretrained HCodec weights are frozen. All audio is resampled to 48 kHz mono, and we randomly crop 4.0-second segments with a per-GPU batch size of 24. We use AdamW with an initial learning rate of \(5\times10^{-4}\), 2000 warm-up steps, and exponential decay. The training loss is a weighted sum across the 16 RVQ layers, with the first layer receiving a weight of 2 and the remaining layers a weight of 1. We apply label smoothing with $\epsilon = 0.1$.

\subsection{Evaluation Metrics}

Due to autoregressive token generation and codec decoding, our outputs are not guaranteed to be sample-aligned with the references. Consequently, sample-level metrics such as scale-invariant signal-to-noise ratio (SI-SNR) and perceptual evaluation of speech quality (PESQ), which require exact sample alignment, are not suitable in our setting \cite{erdogan2023tokensplit}. We therefore adopt perceptual metrics that operate on spectro-temporal representations and are robust to minor temporal shifts.

For overall separation quality, we use ViSQOL (Virtual Speech Quality Objective Listener) \cite{hines2015visqol, chinen2020visqol}, which estimates perceptual audio quality on a scale of 1 to 5. ViSQOL performs internal patch alignment in the spectro-temporal domain, making it tolerant to the small timing variations introduced by autoregressive decoding.

For the vocals track, we additionally evaluate perceptual quality using DNSMOS \cite{reddy2022dnsmos} and NISQA \cite{mittag2021nisqa}. Although both metrics were originally designed for speech, they remain informative for singing vocals, which share core perceptual attributes with speech such as intelligibility, naturalness, and background noise level. DNSMOS reports three sub-scores: signal quality (SIG), background noise quality (BAK), and overall quality (OVRL). NISQA provides a non-intrusive overall quality estimate as an additional proxy for perceived vocal quality.

To remove the influence of codec reconstruction, both the reference signals and the outputs of all baseline models are encoded and decoded through HCodec before evaluation.

\begin{table}[t]
\centering
\setlength{\tabcolsep}{3.5pt}
\small
\caption{Evaluation of source separation performance. Avg denotes the average score across all tracks.}
\label{tab:visqol}
\begin{tabular}{@{}>{\centering\arraybackslash}m{2.0cm} >{\centering\arraybackslash}m{0.6cm} *{5}{c}@{}}
\toprule
\multirow{2}{*}{Model} & \multirow{2}{*}{Type} &
\multicolumn{5}{c}{ViSQOL($\uparrow$)} \\
\cmidrule(l){3-7}
& & vocals & drums & bass & other & avg \\
\midrule
HTDemucs4   & D & 3.72 & 3.88 & 4.11 & 3.11 & 3.71 \\
BS-RoFormer & D & 3.72 & 3.87 & 4.12 & 3.13 & 3.71 \\
SCNet       & D & 3.60 & 3.77 & 3.92 & 3.19 & 3.62 \\
Ours        & G & 3.55 & 3.44 & 4.11 & 3.11 & 3.55 \\
\bottomrule
\end{tabular}
\end{table}

\begin{table}[t]
\centering
\caption{Evaluation of separated vocals track quality.}
\label{tab:vocals}
\begin{tabular}{@{}cccccc@{}}
\toprule
Model       & Type & SIG  & BAK  & OVRL & NISQA \\ \midrule
HTDemucs4 & D    & 2.71 & 3.22 & 2.25 & 2.19  \\
BS-RoFormer & D    & 2.88 & 3.41 & 2.40 & 2.47  \\
SCNet       & D    & 2.65 & 2.89 & 2.17 & 2.33  \\
Ours        & G    & 2.62 & 3.02 & 2.19 & 2.50  \\ \bottomrule
\end{tabular}
\end{table}

\begin{table}[t]
\centering
\caption{Ablation results on source separation quality measured by ViSQOL. Avg is the mean over four tracks.}
\label{tab:ablation}
\begin{tabular}{llccccc}
\toprule
\multirow{2}{*}{Variant} & \multirow{2}{*}{ID} & \multicolumn{5}{c}{ViSQOL($\uparrow$)} \\
\cmidrule(lr){3-7}
 &  & vocals & drums & bass & other & avg \\
\midrule
Main         & -- & 3.55 & 3.44 & 4.11 & 3.11 & 3.55 \\
HuBERT       & A1 & 3.35 & 3.06 & 4.08 & 2.98 & 3.37 \\
Loss weight  & A2 & 3.54 & 3.50 & 4.08 & 3.10 & 3.56 \\
Parallel     & A3 & 3.39 & 3.51 & 4.06 & 3.01 & 3.49   \\
\bottomrule
\end{tabular}
\end{table}

\section{Results}

We compare our generative approach (denoted "G") against three discriminative baselines (denoted "D"): HTDemucs \cite{rouard2023hybrid}, a hybrid time-frequency model; BS-RoFormer \cite{lu2024music} and SCNet \cite{tong2024scnet}, both frequency-domain mask estimation methods.

\subsection{Overall Separation Quality}

Table~\ref{tab:visqol} presents the ViSQOL scores for all four tracks. Our method achieves an average ViSQOL of 3.55, approaching SCNet (3.62), HTDemucs (3.71), and BS-RoFormer (3.71). On the bass track, our model scores 4.11, matching HTDemucs (4.11) and BS-RoFormer (4.12) while outperforming SCNet (3.92). For the other track, our model achieves 3.11, on par with HTDemucs and BS-RoFormer. The remaining gap is mainly observed on drums (3.44 vs.\ 3.77 to 3.88 for the discriminative baselines), suggesting that percussive sources with sharp transients remain challenging for the autoregressive generation paradigm.

\subsection{Vocals Track Quality}
Table~\ref{tab:vocals} reports the DNSMOS and NISQA scores for the separated vocals track. Our method achieves the highest NISQA score of 2.50, surpassing BS-RoFormer (2.47), SCNet (2.33), and HTDemucs (2.19). On the DNSMOS sub-scores, our model obtains a BAK score of 3.02, higher than SCNet (2.89) but lower than HTDemucs (3.22) and BS-RoFormer (3.41). The SIG and OVRL scores are comparable to SCNet, slightly below HTDemucs, and lower than BS-RoFormer.

Figure~\ref{fig:compare} shows a qualitative spectrogram comparison for the vocals track. Compared with BS-RoFormer and SCNet, our method produces a cleaner vocal spectrogram that better matches the target, with fewer residual harmonic structures and transient components attributable to accompaniment leakage. This indicates improved suppression of non-vocal interference and reduced cross-source contamination in the separated vocals.

\subsection{Ablation Studies}

We conduct three ablations to validate key design choices. All variants share the same HCodec, training data, optimizer schedule, and greedy decoding. Results are shown in Table~\ref{tab:ablation}.

\textbf{Conditional encoder (A1).}
Replacing the log-Mel Conformer with a frozen HuBERT extractor drops the average ViSQOL from 3.55 to 3.37, with the largest degradation on drums (3.44 $\rightarrow$ 3.06). This confirms that the learnable Conformer captures mixture-level cues more effectively than frozen speech representations.

\textbf{RVQ loss weighting (A2).}
Changing the layer-wise loss weights from $[2,1,\dots,1]$ to a steeper schedule $[8,4,3,2,2,2,2,2,1,\dots,1]$ yields comparable scores. Notably, it improves the drums score while slightly reducing vocals, bass, and other.

\textbf{Parallel generation (A3).}
Replacing sequential generation with parallel decoding via track-specific output heads lowers the average to 3.49. Vocals and other degrade notably (3.39 and 3.01), suggesting that cross-track autoregressive context benefits separation quality.

\section{Discussion}

\subsection{Limitations}
Our approach faces several limitations. First, the autoregressive generation paradigm struggles with percussive sources characterized by sharp transients, as evidenced by the performance gap on the drums track (3.44 vs.\ 3.77 to 3.88 for discriminative baselines). This may be due to the sequential nature of token-by-token generation, which makes it challenging to precisely model rapid temporal changes. Second, our framework relies on pseudo-labels generated by BS-RoFormer, which may introduce biases and limit the upper bound of performance. Finally, the dual-path codec with 16-layer RVQ requires predicting 32 codebook indices per frame, and cumulative errors across layers can degrade reconstruction quality.

\subsection{Error Analysis}
We conduct qualitative analysis on separation failures across different tracks. Common failure modes include:

\textbf{Vocals:} Bleeding from harmonic instruments (e.g., guitar, piano) in the ``other'' track, especially during sustained notes. Loss of delicate high-frequency details in vocal harmonics. Occasional temporal misalignment in fast passages.

\textbf{Drums:} Blurred attack transients, particularly for snare and kick drums. Loss of cymbal clarity and high-frequency decay. Incomplete separation of hi-hat patterns from other percussive elements.

\textbf{Bass:} Generally good performance, but occasional leakage from kick drum due to overlapping frequency content. Slight loss of low-end definition in complex mixes.

\textbf{Other:} Difficulty distinguishing between similar instruments when they occupy similar frequency ranges. Loss of subtle background textures and ambience.

\section{Conclusion}

We presented a conditional discrete-generation framework for multi-track music source separation that reformulates the task as autoregressive token prediction. Experiments on MUSDB18-HQ validate that this generative paradigm can approach the perceptual quality of established discriminative methods, and even surpass them on the vocals track as measured by NISQA. Ablation studies confirm the effectiveness of the learnable Conformer encoder over frozen speech representations and the benefit of sequential cross-track generation over parallel decoding. However, challenges remain for percussive sources with sharp transients, and the reliance on pseudo-labels may limit the upper bound of performance. Future work will explore dedicated codec designs and specialized tokenization schemes for transient-rich signals to address the performance gap on percussive sources. Additionally, extending the framework to text-conditioned separation and spatial audio upmixing would further enhance its practical applicability.

\section{Generative AI Use Disclosure}

Generative AI tools were used to assist with English language editing and proofreading of this manuscript. All technical content, experimental design, and scientific conclusions are solely the work of the authors.

\bibliographystyle{IEEEtran}
\bibliography{myref}

\end{document}